\documentstyle[12pt,epsfig]{article}          
\catcode`\@=11
\@addtoreset{equation}{section}

        \newdimen\eqskip
        \newdimen\txtskip
        \baselineskip=\txtskip
\def    \be             {\begin{equation}}
\def    \ee             {\end{equation}}
\def    \ba             {\begin{eqnarray}}
\def    \ea             {\end{eqnarray}}

\def    \=              {\;=\;}
\def    \frac           #1#2{{#1 \over #2}}

\def \as   {\mbox{$\alpha_s$}}

\def \oacube {\mbox{${\cal O}(\alpha_s^3)$}}
\def \oatwo {\mbox{${\cal O}(\alpha_s^2)$}}

\def \ppbar {\mbox{$p \bar p$}}
\def \ttbar {\mbox{$t \bar t$}}
\def \pt   {\mbox{$p_{\scriptscriptstyle T}$}}
\def \kt   {\mbox{$k_{\scriptscriptstyle T}$}}
\def \mt   {\mbox{$m_t$}}

\def \ptpair {\mbox{$P^{\ttbar}_{\scriptscriptstyle T}$}}
\def \mpair {\mbox{$M(\ttbar)$}}
\def \dphi  {\mbox{$\Delta\phi$}}
\def \rap   {\mbox{$\eta$}}
\def \mur  {\mbox{$\mu_{\rm \scriptscriptstyle{R}}$}}
\def \muf  {\mbox{$\mu_{\rm \scriptscriptstyle{F}}$}}
\def \muo  {\mbox{$\mu_0$}}
\def \to   {\mbox{$\rightarrow$}}

\begin{document}
\begin{titlepage}
\nopagebreak
{\flushright{
        \begin{minipage}{4cm}
        CERN-TH/95-52 \\
        GeF-TH-3/1995
        \end{minipage}        }

}
\vfill
\begin{center}
{\LARGE { \bf \sc   Top Quark Distributions \\
           in Hadronic Collisions}}
\vfill
{\bf Stefano FRIXIONE}\footnote{Address after June 1: ETH, Z\"urich,
Switzerland}
\vskip .3cm
{INFN, Sezione di Genova, Italy} \\
\vskip .5cm
{\bf Michelangelo L. MANGANO,
\footnote{On leave of absence from INFN, Pisa, Italy}
Paolo NASON,
\footnote{On leave of absence from INFN, Milano, Italy}
\\ \vskip 0.3cm Giovanni RIDOLFI}
\footnote{On leave of absence from INFN, Genova, Italy}
\vskip .3cm
{CERN, TH Division, Geneva, Switzerland} \\
\vskip .6cm
\end{center}
\nopagebreak
\vfill
\begin{abstract}
We present kinematical distributions for top quark pairs produced at
the Tevatron \ppbar\ Collider, as predicted within Next-to-Leading-Order QCD.
We consider single and double-inclusive distributions, and compare our results
to those obtained with the shower Monte Carlo HERWIG. We discuss the
implications of our findings for experimental issues such as the measurement of
the top quark mass.
\end{abstract}
\vskip 1cm
CERN-TH/95-52 \hfill \\
March 1995 \hfill
\vfill
\end{titlepage}
Now that the existence of the $top$ quark has been firmly established
via its detection in hadronic collisions \cite{cdf2,cdf1,d0}, experimental
studies will focus on the determination of its properties. In particular, the
measurement of its mass and of the production cross section and distributions
will certainly be among the first studies of interest. The top mass represents
one of the crucial parameters for testing the Standard Model \cite{ewk-tests},
while the production properties, should they display anomalies, could point to
the existence of exotic phenomena \cite{lane-et-al}.

Early estimates of the total production cross section for top quarks
\cite{sigtot,gual}\ have later been updated \cite{kellis}.
In ref. \cite{sigres} a
partial resummation of leading logarithms from soft gluon emission
has also been included.

In this Letter we present a set of kinematical distributions of relevance for
the study of production dynamics, as obtained from full Next-to-Leading-Order
(NLO) QCD perturbative calculations.  We will consider inclusive \pt\ and
pseudorapidity (\rap) distributions, as well as invariant mass (\mpair),
transverse momentum (\ptpair) and azimuthal correlations (\dphi) of the top
quark pair.

The inclusive \pt\ distribution is sensitive to channels such as $W g \to t
\bar b$ \cite{wg}, which are found to contribute with a small cross section,
predominantly at low \pt.
The invariant mass of the pair is an obvious probe of the possible existence of
strongly coupled exotic resonances, such as technimesons \cite{lane-et-al}.
The transverse momentum of the pair is an indication of
the emission of hard hadronic jets in addition to those coming from the top
decays. The presence of these additional jets generates potentially large
combinatorial backgrounds to the reconstruction of the top mass peak
from the decay products \cite{cdf1}.
An accurate understanding of these backgrounds is in principle
very important for a precise measurement of the top mass.

The NLO total, single and double inclusive cross sections for the production of
heavy quarks in hadronic collisions have been calculated in references
\cite{sigtot}, \cite{pt} and \cite{mnr}, respectively.
Since most experimental studies are performed using shower Monte Carlo event
generators \cite{montecarlo}\ to simulate the behaviour of the events in the
detectors, we will compare our NLO results with those we obtained with HERWIG
\cite{herwig}. This is important in order to assess the reliability of the
theoretical inputs  used by the experiments. It has been recently pointed out
\cite{orr}\ that there are possibly significant differences in the jet activity
predicted by fixed-order parton level calculations and the results of HERWIG,
with possible implications for the experimental mass measurement. We will
discuss here a possible origin of these discrepancies.

For our calculations we use \mt=176 GeV (CDF, ref. \cite{cdf2}).
Differences of $\pm$ 15 GeV, approximately the current experimental error,
do not affect the shapes of the distributions we present nor any of our
conclusions. Our default set of parton densities is MRSA \cite{mrsa}, and we
explore a range of renormalization and factorization scales $\muo/2 < \mur=\muf
< 2 \muo$, with $\muo^2 = \mt^2+\pt^2$  ($\muo^2 = \mt^2+((\pt)_t^2+(\pt)_{\bar
t}^2)/2$) for single (double) differential distributions.

Throughout our plots, we rescale the HERWIG calculations by the perturbative
$K$ factor given by the ratio of the \oacube\ and the \oatwo\ results. The $K$
factor is of the order of 1.3 for all choices of parameters.
The distributions of \pt, \rap\ and \mpair\ are shown in
figures 1--3.
\begin{figure}[htb]
\centerline{\psfig{figure=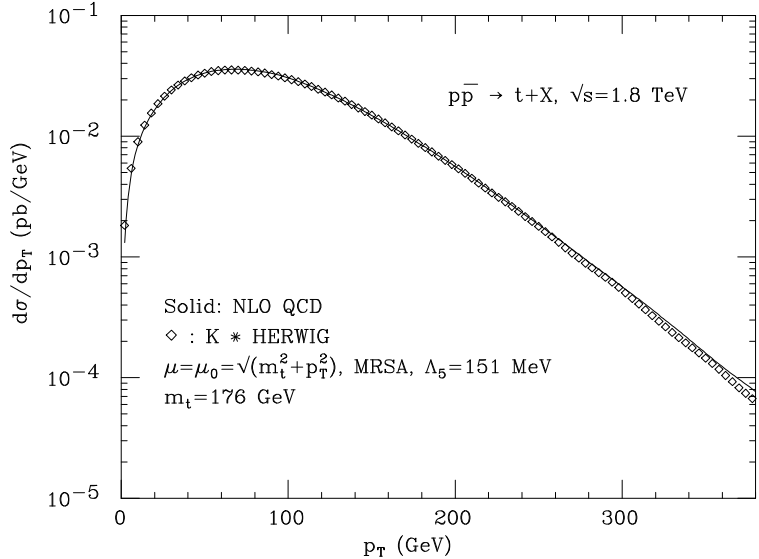,width=10cm,clip=}}
\caption{ \label{ptinc}
Inclusive \pt\ distribution of the top quark.
}
\end{figure}
\begin{figure}[htb]
\centerline{\psfig{figure=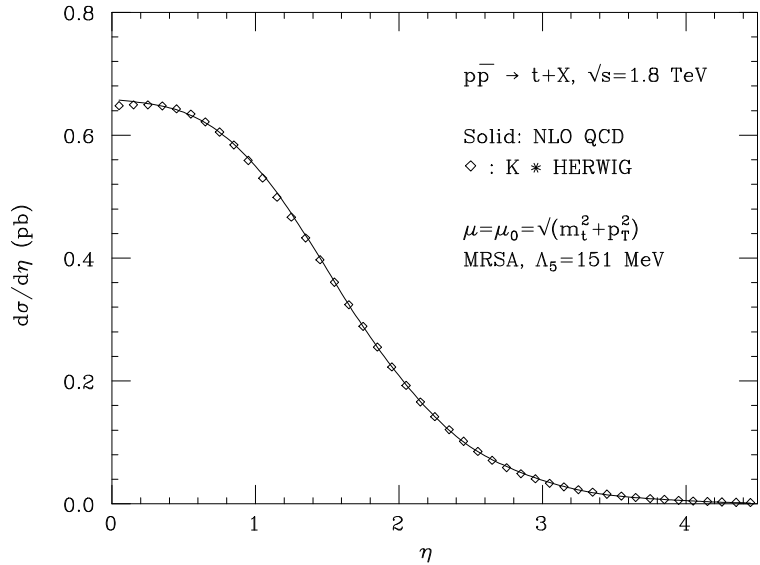,width=10cm,clip=}}
\caption{ \label{etainc}
Inclusive pseudorapidity distribution of the top quark.
}
\end{figure}
\begin{figure}[htb]
\centerline{\psfig{figure=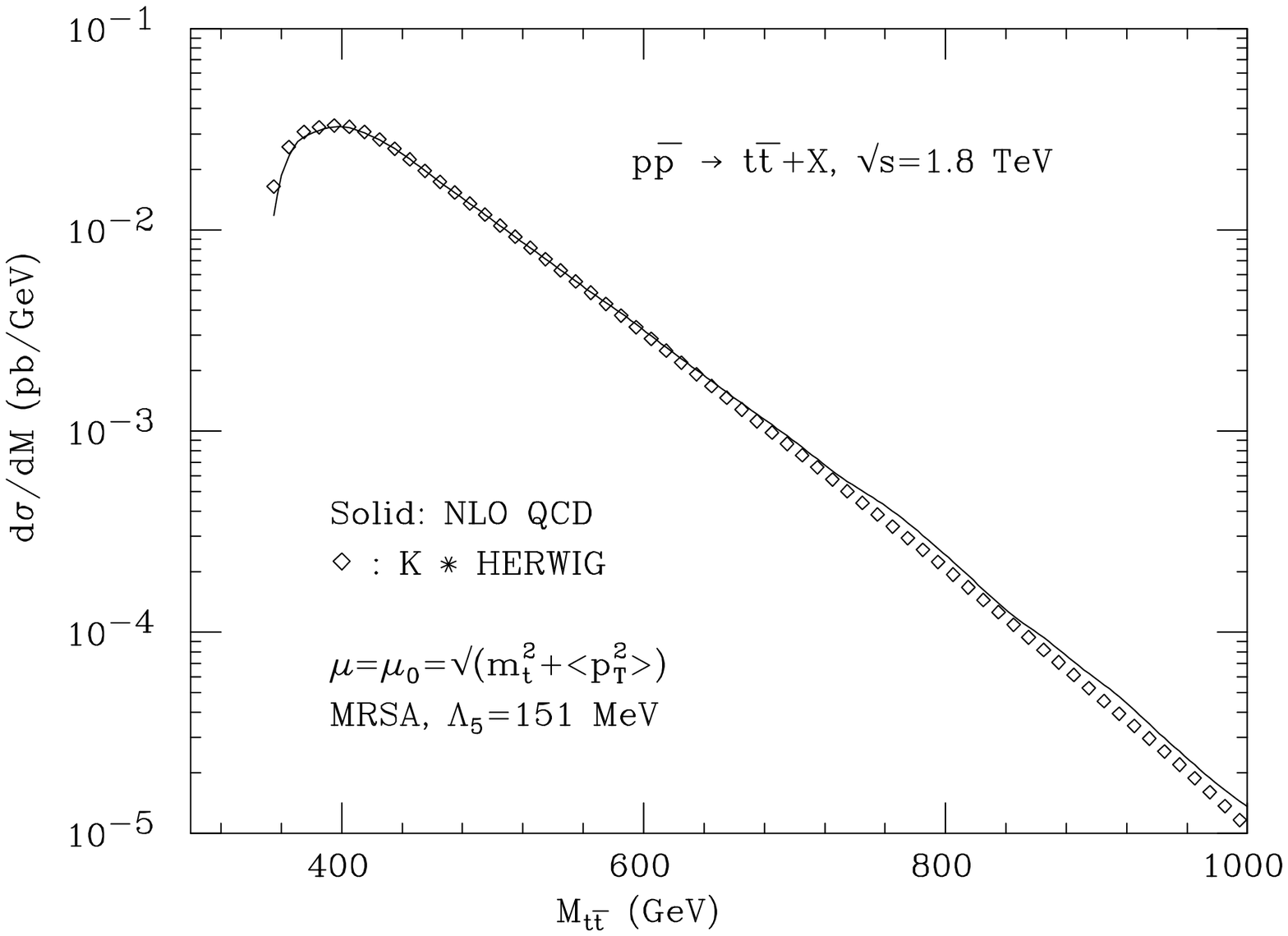,width=10cm,clip=}}
\caption{ \label{mass}
Invariant mass distribution of the $t\bar{t}$ pair.
}
\end{figure}
The solid lines correspond to the NLO results obtained
using \mur=\muf=\muo. The curves obtained using a different choice of scales,
within the range quoted earlier, result in an overall normalization change, by
approximately $\pm 10\%$. No change in the shapes is observed, and we therefore
decided not to include these curves in the plots for the sake of clarity.
The square points correspond to the HERWIG prediction for \mur=\muo, rescaled
by a $K$ factor equal to 1.34.

The figures indicate clearly that distributions which are not trivial at
leading order, such as \pt, \rap\ and \mpair, agree perfectly between the
NLO and HERWIG calculations, at least for values of \pt\ and \mpair\ not too
large. This result follows naturally both from the observed
stability of those distributions under radiative corrections regardless of the
heavy quark mass \cite{mnr}, and from the large inertia of the top
quark against the combination of higher order and non-perturbative
corrections introduced by the Monte Carlo.
At large \pt\ and \mpair, however, one starts observing a small
deviation, due to the
multiple gluon emission from the final state top quarks. This effect
is not described by the NLO QCD calculation,
which only includes one gluon emission.
Multiple gluon emission becomes relevant for $\pt\gg\mt$, where it
leads to a top quark fragmentation function softer than predicted by the fixed
order NLO calculation.
Due to the smallness of the cross section in these regions, large statistics
will be needed in order to test this behaviour.

Those distributions which are trivial at leading order,
\dphi\ and \ptpair, are viceversa most
sensitive to multiple gluon emission from the initial state.
This is because even small perturbations
can smear a distribution which at leading order is represented by a delta
function, which is the case of \ptpair\ and \dphi. The most visible effect is
observed in \ptpair, Figure~4, where
\begin{figure}[htb]
\centerline{\psfig{figure=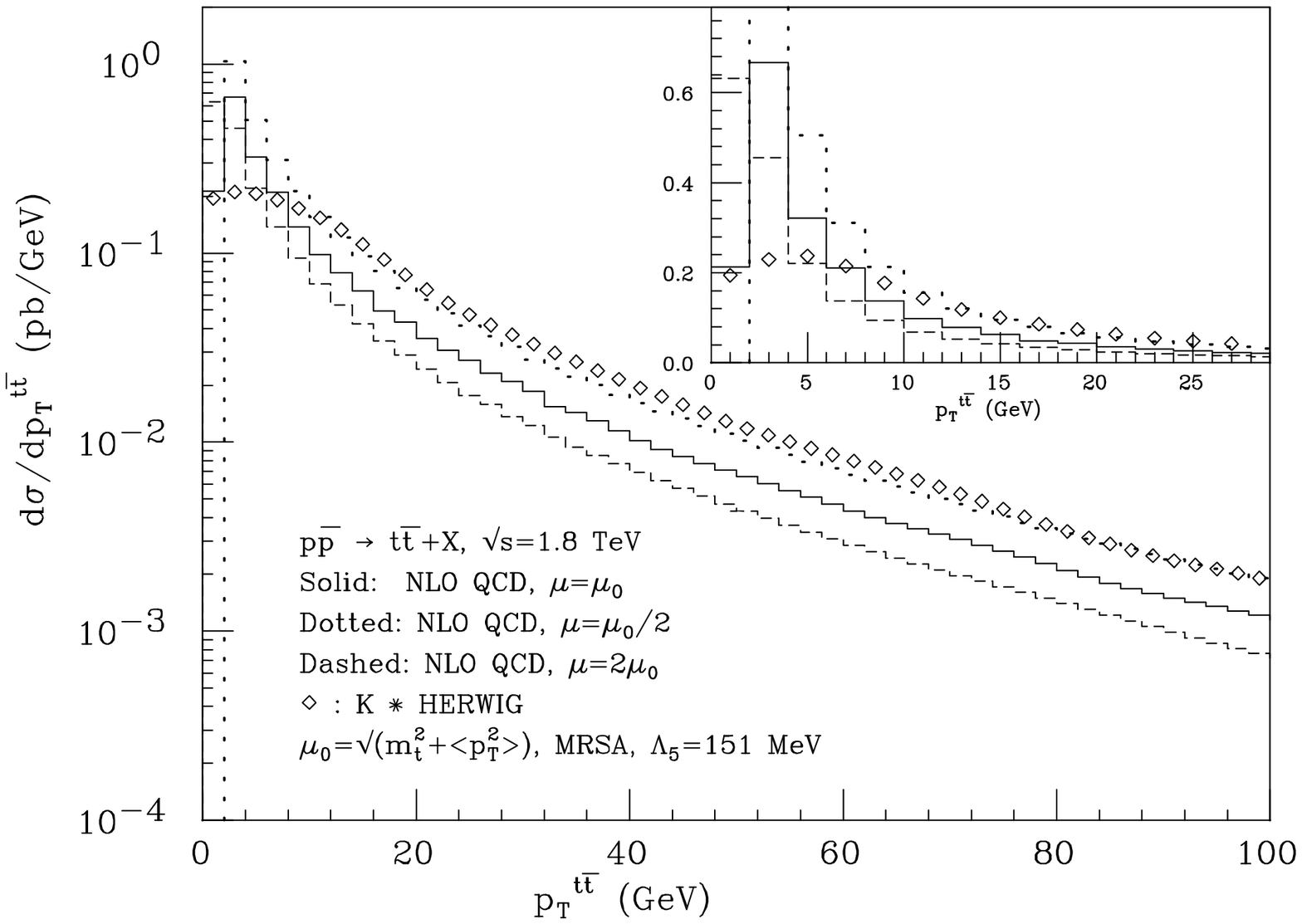,width=13cm,clip=}}
\caption{ \label{ptpair}
Transverse momentum distribution of the $t\bar{t}$ pair.
}
\end{figure}
we include the NLO curves relative to the
three choices of scales, \mur=\muf=\muo (solid), \muo/2 (dots) and 2\muo
(dashes). Contrary to the previous cases, significant differences in shape
arise here among the three choices in the small \pt\ region.
The HERWIG result (normalized to the area of the solid curve) is also shown.
The NLO and the HERWIG distributions assume the same shape only
for \ptpair\ larger than approximately 20 GeV.

To better quantify the differences, we quote in
Table~1 the value of the average \ptpair\ and of the fraction of the total
cross section with $\ptpair>10,\,20\;$GeV for the different curves appearing on
the plot.
{\renewcommand{\arraystretch}{1.8}
\begin{table}
\begin{center}
\begin{tabular}{lccc} \hline\hline
  & $\langle \ptpair \rangle$ (GeV) &  F($\ptpair>10$ GeV) (\%)
  &  F($\ptpair>20$ GeV) (\%) \\ \hline\hline
NLO QCD, \mur=\muf=\muo   &  11.9 &    30 & 15 \\ \hline
NLO QCD, \mur=\muf=2\muo  &  9.1  &    23 & 11 \\ \hline
NLO QCD, \mur=\muf=\muo/2 &  16.6 &    44 & 22 \\ \hline
HERWIG,  \mur=\muf=\muo   &  17.5 &    51 & 28 \\ \hline
\end{tabular}
\caption{Average value of \ptpair\ and fraction of the cross section with
$\ptpair>10$ and 20 GeV for different calculations.
 }
\end{center}
\end{table} }
First of all we notice the strong scale dependence of the results in
Table~1. This is a first indication of the lack of reliability of the
fixed order calculation for these particular quantities.
Observe that the NLO result obtained with the lowest scale is closer
to the HERWIG result. In fact, a lower scale is presumably more appropriate
in this problem. This is because there are two relevant
scales in this case: the mass of the top, and the transverse
momentum cut. Since the production of top quarks takes place
mainly through the exchange of a highly virtual gluon (with virtuality
of the order of the top mass), while the emission of the jet
takes place at the smaller scale of the jet transverse momentum,
it is sensible to take in the \oacube\ cross section
two powers of $\as$ at the large scale $\muo$
and one at the transverse momentum of the jet.
We have performed this exercise, and obtained
$\sigma(\ptpair>10\;\mbox{GeV})=1.987\,{\rm pb}$,
which is 43\% of the total cross section computed at the scale $\muo$.

The high value of $\langle \ptpair \rangle$ for the NLO calculation indicates
that the probability of emitting a gluon with transverse energies \kt\ of the
order of 10 GeV is of order 1. Technically, this is related to the appearance
of a large double logarithm ${\cal{O}} (\as \log^2[m_t^2 / k_T^2])$.
It is well known \cite{sudakov}\ that in these circumstances multiple
gluon emission, corresponding to multiple powers of these double logs,
needs to be resummed. A typical example can be found in $W$ and $Z$ production
at small \kt \cite{wzresum}.

The Monte Carlo HERWIG performs the resummation of leading and some
subleading double logarithms,
faithfully reproducing the behaviour of the analytically resummed
cross sections in known cases \cite{herwig}.
The large top mass extends the region where
Sudakov effects are important up to \kt\ values which are often considered
within the domain of applicability of standard fixed-order
perturbation theory.

We have implemented a leading logarithm version of the resummation procedure
described by Collins Soper and Sterman in \cite{wzresum}, and qualitatively
reproduced  the behaviour of the
Monte Carlo\footnote{A similar calculation
in the context of $b$ pair production has been given in ref.~\cite{berger2}.}.
We therefore conclude that the parton level result (with
the factorization and renormalization scale taken of the order of the
mass of the top quark)
is less reliable than the Monte Carlo in the small \ptpair\ region,
and gives a cross section (for \ptpair$>10\;$GeV) which
is roughly 60\% of what one would get with a more judicious choice of
the scale, and with the inclusion of Sudakov effects.
We also point out that no choice of scales in the NLO calculation
would describe the detailed shape in the region $\ptpair<20\;$GeV.
On the other hand, we expect the parton calculation to be more
accurate in the large \ptpair\ region, where the soft
approximation used in the Monte Carlo breaks down.

A similar problem should be expected in the description
of jet radiation during the production of any massive system, in particular
W or Z boson pairs. This might affect parton level calculations of backgrounds
to \ttbar\ production in the dilepton plus two jet channel.
A simple estimate along the lines of what performed above
for the top, sets the threshold below which fixed order perturbation theory
becomes unreliable at about 10-15 GeV, which still corresponds to jets
sufficiently energetic to be reconstructed as such by the Collider
detectors.
A confirmation of this estimate can be found in the study of resummation
effects on the \pt\ distribution of Z pairs presented in reference~\cite{han}.

The observations made above partially
explain the disagreement found in ref. \cite{orr} between the
relative fraction of jets emitted during the production process and jets
emitted from the $b$ quark resulting from the top decay. These authors found
that their parton level calculation gave
rise to a ratio smaller than HERWIG's. They interpreted it as a deficit in jet
radiation off the final state $b$ quark within HERWIG.
In order to make a more definite comparison with the results of
ref. \cite{orr}, we have computed the
$t\bar{t}\;{\rm jet}$ cross section with the same value of the
parameters and the same set of cuts that they applied.
We used therefore $m_t=174\;$GeV, $\muo=m_t$, $|\eta_{\rm jet}|<2.5$,
$\ptpair>10\;$GeV. In order to emulate the cuts they perform on the
$b$ jets coming from top decays, we let the top quark decay
isotropically, and imposed the additional cuts $|\eta_{\rm b}|<2.5$
and $p_{\scriptscriptstyle T}^{(b)}>10\;$GeV. We verified that these additional
cuts affect the cross section by less than 3\%.
We obtained a cross section
of 1.30 pb, to be compared with the value of 1.0 pb of ref. \cite{orr}.
We have no explanation for this difference.
{}From the considerations made previously, we expect that the inclusion
of Sudakov effects should increase the cross section
by roughly 66\%, leading to a value of 2.16 pb,
more than a factor of two larger than the result of ref.~\cite{orr}.

Figure~5, finally, shows the \ttbar\ azimuthal correlations.
\begin{figure}[htb]
\centerline{\psfig{figure=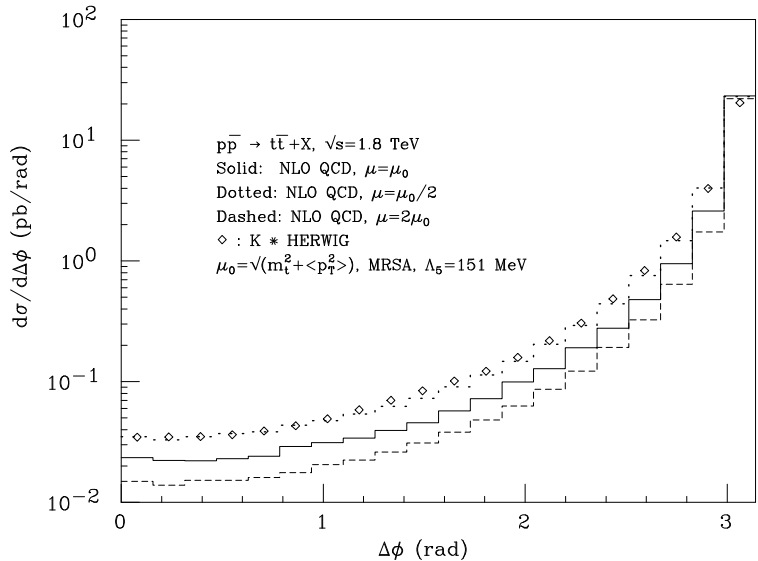,width=10cm,clip=}}
\caption{ \label{dphi}
Azimuthal distance distribution of the $t\bar{t}$ pair.
}
\end{figure}
Notice that
most of the shape discrepancies observed at \ptpair$<$20 GeV are forced here
into the bin around the \dphi=$\pi$ point. This
is because even a 20 GeV kick will only deflect the trajectory of the
$t$ or $\bar{t}$ quark (each traveling with momentum of the order of
\mt/2), by about 1/10 of a radian.

While completing this paper, a study of single inclusive distributions of top
quarks at the Tevatron appeared \cite{smith}.
These authors improved the NLO calculation with the inclusion of a partial
resummation of leading soft logarithms \cite{sigres}. Their results indicate
that single inclusive distributions are only affected in overall normalization,
but not in shape (at least for \pt\ not much larger than \mt). This is
consistent with the results presented here, and provides yet more
confidence in the validity of the perturbative predictions applied
to single inclusive quantities.

In concusion, we presented here a set of kinematic quantities which should be
used to test the dynamics of top quark production in hadronic collisions.
Most of these quantities were shown to be subject to very small theoretical
uncertainty within NLO perturbative QCD, and to be rather insensitive to higher
order or hadronization effects. One noticeable exception is the
transverse momentum of the pair, which is in turn related to the spectrum of
jets recoiling against the top pair. We found in our study that an accurate
description of the region \ptpair$<20$ GeV requires the resummation of leading
soft and collinear double logarithms, as implemented in appropriate shower
Monte Carlo programs.


\begin{thebibliography}{99}
\def    \nuke   #1#2#3{{\sl Nucl. Phys.} {\bf B#1}  (#2) #3}
\def    \pl     #1#2#3{{\sl Phys. Lett.} {\bf #1B}  (#2) #3}
\def    \prl    #1#2#3{{\sl Phys. Rev. Lett.} {\bf #1}  (#2) #3}
\def    \pr     #1#2#3{{\sl Phys. Rev.} {\bf #1}  (#2) #3}
\def    \prd    #1#2#3{{\sl Phys. Rev.} {\bf D#1}  (#2) #3}
\def    \prep   #1#2#3{{\sl Phys. Rep.} {\bf #1}  (#2) #3}
\bibitem{cdf2}
        F. Abe et al., CDF Collab., Fermilab-Pub-95-022-E.,
        submitted to Phys. Rev. Lett., Febr~24~1995;\\
        S. Abachi et al., D0 Collab., submitted to Phys.
        Rev. Lett., Febr~24~1995.
\bibitem{cdf1}
        F. Abe et al., CDF Collab., \prd{50}{1994}{2966}.
\bibitem{d0}
        S. Abachi et al., D0 Collab., FERMILAB-PUB-94-354-E, to appear on
        Phys. Rev. Lett.
\bibitem{ewk-tests}
        The LEP Collaborations, ALEPH, DELPHI, L3 and OPAL, and the
        LEP Electroweak Working Group, CERN/PPE/94-187.
\bibitem{lane-et-al}
        E. Eichten and K. Lane, \pl{327}{1994}{129};  \\
        C.T. Hill and S.J. Parke, \prd{49}{1994}{4454}.
\bibitem{sigtot}
        P.~Nason, S.~Dawson and R.~K.~Ellis,
        \nuke{303}{1988}{607};
        W.~Beenakker, H. Kuijf, W.L. van Neerven and J. Smith,
        \prd{40}{1989}{54}.
\bibitem{gual}
        G. Altarelli, M. Diemoz, G. Martinelli and P. Nason,
        \nuke{308}{1988}{724}.
\bibitem{kellis}
        R.K. Ellis, \pl{259}{1991}{492}.
\bibitem{sigres}
        E. Laenen, J. Smith and W.L. van Neerven,
        \nuke{369}{1992}{543}; \pl{321}{1994}{254}.
\bibitem{wg}
        S. Dawson, \nuke{284}{1985}{449};\\
        S. Willenbrock and D.A. Dicus, \prd{34}{1986}{155};\\
        C.P. Yuan, \prd{41}{1990}{42};\\
        R.K. Ellis and S.J. Parke, \prd{46}{1992}{3785}; \\
        G. Bordes and B. van Eijk, \nuke{435}{1995}{23}.
\bibitem{pt}
        P.~Nason, S.~Dawson and R.~K.~Ellis,
        \nuke{327}{1988}{49};
        W.~Beenakker et al.,
        \nuke{351}{1991}{507}.
\bibitem{mnr}
        M. Mangano, P. Nason and G. Ridolfi,
        \nuke{373}{1992}{295}.
\bibitem{montecarlo}
        B.R. Webber, {\em Annu. Rev. Nucl. Part. Sci.}
        {\bf 36}(1986)253;\\
        R. Odorico, {\it Computer Phys. Comm.} {\bf 32}(1984)139;  \\
        G. Marchesini and B.R. Webber, \nuke{238}{1984}{1};\\
        B.R. Webber, \nuke{238}{1984}{492};\\
        H.-U. Bengtsson and G. Ingelman, {\it Computer Phys. Comm.}
        {\bf 34}(1985)251;\\
        R.D. Field, \nuke{264}{1986}{687};\\
        F. Paige and S.D. Protopopescu, Brookhaven report BNL-38034 (1986);\\
        T. Sj\"{o}strand and M. Bengtsson, {\it Computer Phys. Comm.}
        {\bf 43}(1987)367;\\
        B. Andersson, G. Gustafson and B. Nilsson-Almqvist,
        \nuke{281}{1987}{289}.
\bibitem{herwig}
        G. Marchesini and B.R. Webber, \nuke{310}{1988}{461}. In this work we
        used version 5.8 of the program, after modifying the default choice of
        scale for the hard process to \mur=\muf=\muo.
\bibitem{orr}
        L.H. Orr, T. Stelzer and W.J. Stirling, DTP/94/112, hep-ph/9412294.
\bibitem{mrsa}
        A.D. Martin, R.G. Roberts and W.J. Stirling, \prd{50}{1994}{6734}.
\bibitem{sudakov}
        Yu.L. Dokshitzer, D.I. Dyakonov and S.I. Troyan,
        \prep{58}{1980}{271};\\
        G. Parisi and R. Petronzio, \nuke{154}{1979}{427}.
\bibitem{wzresum}
        G. Altarelli, R.K. Ellis, M. Greco and G. Martinelli,
        \nuke{246}{1984}{12};\\
        J.C. Collins, D.E. Soper and G. Sterman, \nuke{250}{1985}{199}.
\bibitem{berger2}
        E. Berger and R. Meng,
        \prd{49}{1994}{3248}.
\bibitem{han}
        T. Han, R. Meng and J. Ohnemus, \nuke{384}{1992}{59}.
\bibitem{smith}
        N. Kidonakis and J. Smith, ITP-SB-94-63, hep-ph/9502341.
\end{thebibliography}
\end{document}